\documentclass[journal]{IEEEtran}

\ifCLASSINFOpdf
\else
   \usepackage[dvips]{graphicx}
\fi
\usepackage{url}
\usepackage{braket}
\usepackage{amsmath, amssymb}
\usepackage{amsthm}
\usepackage{graphicx}
\usepackage{tabularx}
\usepackage{array, boldline, rotating}

\newcommand{\simPG}[1][x]{\sim \frac{\mathcal{P}(a #1)}{a} + \mathcal{N}(0,b^2)}
\newcommand{\card}[1]{\left| #1 \right|}

\newcommand{\thickhline}{\hlineB{4}}

\usepackage[T1]{fontenc}
\usepackage[pdfencoding=auto]{hyperref}

\begin{document}

\title{Target Aware Poisson-Gaussian Noise Parameters Estimation from Noisy Images}

\author{Étienne Objois \hspace{5em} Kaan Okumuş \hspace{5em} Nicolas Bähler\\
    Supervisor: Majed El Helou, Ph.D.\\
    Professor: Prof. Sabine Süsstrunk
}

\date{June 3, 2022}

\markboth{Computational photography, CS-413 - EPFL}
{Shell}
\maketitle

\begin{abstract}
    Digital sensors can lead to noisy results under many circumstances. To be able to remove the undesired noise from images, proper noise modeling and an accurate noise parameter estimation is crucial. In this project, we use a Poisson-Gaussian noise model for the raw-images captured by the sensor, as it fits the physical characteristics of the sensor closely. Moreover, we limit ourselves to the case where observed (noisy), and ground-truth (noise-free) image pairs are available. Using such pairs is beneficial for the noise estimation and is not widely studied in literature. Based on this model, we derive the theoretical maximum likelihood solution, discuss its practical implementation and optimization. Further, we propose two algorithms based on variance and cumulant statistics. Finally, we compare the results of our methods with two different approaches, a CNN we trained ourselves, and another one taken from literature. The comparison between all these methods shows that our algorithms outperform the others in terms of MSE and have good additional properties.
\end{abstract}


\section{Introduction}

Every image capturing system is inherently noisy. The noise is influenced by
different factors and different systems have different noise characteristics. In
our project, we pick a model of noise having two components, one being a Poisson distribution and the other one a Gaussian.

Roughly, capturing an image can be seen as the process of counting the number of
incident photons that hit a sensor pixel during a given amount of time. More
photons in a given interval of time translates to more light and hence more
intensity for the pixel of the final image. Hence, the Poisson distribution is
inherent to that discrete photon counting phenomenon. The Poisson
contribution in that context is commonly referred to as photon shot noise. The second element of our noise model, the Gaussian, is introduced by a
collection of different error factors like the quantum efficiency, the
circuitry, unwanted interactions between pixels, read out errors and many more.
Overall, all those error sources combined can be modeled with a single Gaussian.

Our goal is to estimate the parameters of this noise model. Knowing those values enables performing
noise correction. More precisely, from an observed noisy image, $y$ reconstruct the ground-truth image $x$. In our setting, we assume to have access to
both $y$ and $x$. This assumption is reasonable in a calibration setting where one can do long
exposure times to minimize the Poisson contribution and average over several images
to reduce the impact of the Gaussian noise part. Once calibrated (i.e., having estimated the noise parameters) newly captured images (without knowing the ground-truth) can be corrected for the noise  leading to better results.

Generally, there are two main approaches to noise estimation today, either using
statistical techniques and signal processing or deep learning. The former
involves more domain specific knowledge.

The method presented by Foi et al.~\cite{foi2008practical} follows the ideas of the first approach.
Additionally, it uses a Poisson-Gaussian noise model like we do but only uses
observations of the
noisy signal, not the ground-truth $x$. Hence, the problem the
authors of~\cite{foi2008practical} try to solve is inherently more difficult than ours. In our setting we have knowledge of both $y$ and $x$, hence, this advantage
should enable us to achieve better performance.

On the other hand, deep learning is increasingly often applied to all kinds
of fields, noise estimation is no exception. Specifically, Convolutional Neural
Networks (CNN) that are abundantly used in many image related tasks. Here, we are not limited to using only $y$ but also $x$ and maybe even $|y-x|$

In this project, we propose novel methods of noise estimation while comparing
their performance to different approaches. Further, we put those results into
perspective by providing the log-likelihood we derived for this problem.

In the case where both $x$ and $y$
are at hand, our findings allow improving over the conventional methods.

For our method to work, we heavily rely on the knowledge of the noise-free
ground-truth image $x$. Further, we are only working with grayscale images, but
our methods are extendable to multichannel images. Each channel's noise
parameters might be different from each other, as each channel is independent
from any other. Additionally, we didn't address the issue of clipping,
i.e., handling values that lay outside the range of valid pixel intensities. For
instance, intensity is given by a value in $[0., 1.]$ and any pixels' intensity beyond
this interval should be clipped to it's closest bound of the interval in order
for it a valid value. But
clipping is introducing a nonlinearity which makes all the derivations we make
more complex. For simplicity, we allowed values to exceed the range and do not
apply any clipping.

\section{Related work}\label{sec:related}
Denoising is one of the most fundamental tasks in image restoration, with both
theoretical impact and practical applications. Most classic denoisers, for
instance PURE-LET~\cite{luisier2011image}, KSVD~\cite{KSVD}, WNNM~\cite{WNNM},
BM3D~\cite{BM3D}, and EPLL~\cite{EPLL}, require knowledge of the noise level in
the input test image. Deep learning image denoisers that have shown improved
empirical performance~\cite{huang2022neighbor2neighbor,ma2021deep} also require
knowledge of noise distributions, if not at test time~\cite{ffdnet}, then at
least for training~\cite{elhelou2020blind,elhelou2022bigprior}. This is due to
the degradation overfitting of deep neural networks~\cite{el2020stochastic}.
Noise modeling is thus important for denoisers at test time, but also for
acquisition system analysis and dataset modeling for training these denoisers.
Past research has focused on modeling noise from noisy images without relying on
ground truth, i.e., noise-free, information~\cite{foi2008practical}. Interesting
approaches, for example Sparse Modeling~\cite{7990585}, Dictionary
Learning~\cite{2018} or non-local image denoising methods like
SAFPI~\cite{10.1371/journal.pone.0208503}, have been developed to push overall
denoising performance. However, none of these methods allow easy use of
noise-free data when it is available. For Poisson-Gaussian noise modeling, for
example, both FMD~\cite{zhang2019poisson} and W2S~\cite{zhou2020w2s} rely on a
noise modeling method that does not consider ground truth noise-free
images~\cite{foi2008practical}. Hence, our approach to model the
\textbf{Po}isson-\textbf{Ga}ussian \textbf{I}mage \textbf{N}oise (PoGaIN)
distribution exploits paired samples (noisy and noise-free images), which
significantly improves the modeling accuracy. Our method is based on the
cumulant expansion, which is also used by other authors to derive estimators for
PoGaIN model parameters, but for different input types, such as noisy image time
series~\cite{6235897} or single noisy images~\cite{zhang:pastel-00003273}. We
lastly refer the reader to our concise publication that sums up the essential
elements of this report~\cite{9976220}.

\section{Theory}

\subsection{Poisson-Gaussian Modeling}
The generic signal-dependent Poisson-Gaussian noise modeling can be written as
the following form :
\begin{equation}
    y = \frac{1}{a} \alpha + \beta, \quad \alpha \sim \mathcal{P}(ax), \quad \beta \sim \mathcal{N}(0,b^2)
    \label{eq:poisgaus1}
\end{equation}
where $x$ is the known ground-truth signal and $y$ is the observed signal.
In our modeling, Poisson signal-dependent component $\eta_p$ and Gaussian
signal-independent component $\eta_g$ are defined as,
\begin{equation}
    \eta_p = \frac{1}{a} \alpha, \quad \eta_g = \beta
\end{equation}
where these two components are assumed to be independent.
From the derivation given in Appendix~\ref{appendix:derivML}, the following properties of
the observed signal, $y$, can be found :
\begin{equation}
    \mathbb{E}[\eta_p] = x, \quad \mathbb{V}[\eta_p] = \frac{x}{a}
\end{equation}
Here, the fact that Poisson noise has signal-dependent characteristics. On the other hand, the Gaussian noise has the constant variance and mean, which makes it signal independent as expected.
Consequently, the following equation~\ref{eq:statpoisgaus} is obtained.
\begin{equation}
    \mathbb{E}[y]=x, \quad \mathbb{V}[y] = \frac{x}{a} + b^2
    \label{eq:statpoisgaus}
\end{equation}
Intuitively, it means that the average of the observed image should be the ground-truth image, which justifies the reasoning. From the variance equation, the fact that variance is affected directly by $a$ and $b$ makes it reasonable as they represent the noises.

\subsection{Raw-Data Modeling}
Poisson-Gaussian model is properly matched with the natural characteristics of
raw-data of digital imaging systems. The Poisson noise models the
signal-dependent part of errors, which are caused by the discrete nature of
the photon-counting process. On the other hand, Gaussian noise models the signal
independent errors, such as electric and thermal noise.

The parameter of the Poisson noise, $\alpha$ is dependent on the quantum
efficiency of the sensor. The more the number of photons to generate the
electrons inside the sensor is, the less the value of $\alpha$ is. According
to experiments conducted by~\cite{rawdatamodeling:poissonnoiseeffect}, this
Poisson noise effect can be the dominant contributor to uncertainty in the raw
data captured by high-performance sensors. This justifies the accuracy of the modeling for raw-image of the sensors.

Analog gain which is the amplification of the collected charge in the digital
imaging systems is another dependent that affects both Poisson and Gaussian
noise parameters. In digital camera, it is controlled by ISO and/or exposure
index (EI) sensitivity settings. The larger the ISO number is, the larger analog
gain resulted in, which causes the amplification of the noises as well. This
causes the decrease in SNR of the captured raw-data, which means the noise increases. We can conclude that both of the noise parameters can be highly dependent on the analog gain.

In the case of the system with large photon counting condition, Poisson noise
can be approximated as Gaussian noise as the following.
\begin{equation}
    \mathcal{P}(\lambda) = \mathcal{N}(\lambda, \lambda)
\end{equation}
This approximation can be useful for deriving the solution based only on mean
and variance, which simplifies the process of proposing algorithms without using
ground-truth image~\cite{foi2008practical}. However, since we were also looking for the
method that uses the ground-truth image as input, this approximation is not
applied for the following sections. Another reason is that this approximation results in the loss of information about the statistics of the actual noise parameters. In other words, it results in lossy projection of $a$ and $b$ into less dimensional space, which is not desirable.

\subsection{Maximum Likelihood Solution}
When the noise modeling in equation~\ref{eq:poisgaus1} is applied to a raw-data image, the
following likelihood function of the pixel intensity of an observed image can be
achieved with the derivation explained in Appendix~\ref{appendix:derivML}.
\begin{equation}
    f_y(y_{n}|a,b,x) =\sum_{k=0}^\infty \frac{(ax_n)^k}{k!b\sqrt{2\pi}}\exp{\left(-ax_n-\frac{(y_n-k/a)^2}{2b^2}\right)}
    \label{eq:likelihood_func}
\end{equation}
where $y$ is observed image, $x$ is the ground-truth image and $n$ is the pixel index.

In order to propose a robust noise parameter estimation algorithm, the optimality
criterion is chosen to be the maximization of the likelihood function in
~\ref{eq:likelihood_func} with respect to noise parameters, $a$ and $b$. The
resulted solution of this optimality criterion is called as Maximum Likelihood
solution. From the derivation in~\ref{appendix:derivML}, the following solution
is found.
\begin{equation}
    \begin{split}
        \hat{a}, \hat{b} = \arg \max_{a, b} \prod_{n}\sum_{k=0}^\infty \frac{(ax_n)^k}{k!b\sqrt{2\pi}}\exp{\left(-ax_n-\frac{(y_n-k/a)^2}{2b^2}\right)}
    \end{split}
    \label{mlsoln}
\end{equation}

\section{Implemented Methods}

\subsection{Grid Search for ML Solution}
Maximum Likelihood solution offers an accurate estimation of the noise parameters in theory. However, for the practical reasons, it's hard to propose the algorithmic solution for the maximization of the functional inside the ML solution. This functional in equation~\ref{mlsoln} is analyzed and found to be non-concave. Thus, gradient-based optimization algorithms cannot be applied for this maximization problem. For the sake of implementation of ML solution, the most naive method is proposed to estimate the noise parameters $a$ and $b$. This is also possible, as we only have two parameters to be estimated.

As an implementation issue, exact calculation of likelihood function is difficult as it includes infinity sum as seen in equation~\ref{eq:likelihood_func}. In order to approximately estimate it, a sufficiently large value of $k_{max}$ is chosen, and the following is applied:
\begin{equation}
    f_y(y_{n}|a,b,x) \approx \sum_{k=0}^{k_{max}} \frac{(ax_n)^k}{k!b\sqrt{2\pi}}\exp{\left(-ax_n-\frac{(y_n-k/a)^2}{2b^2}\right)}
    \label{eq:kmax_likelihood_func}
\end{equation}

However, from the analysis of the non-concavity behavior of the likelihood function, it's found that it does not result in sufficiently good results for the estimation of noise parameters. In order to achieve the accurate results, very small step sizes should be chosen. This makes the algorithm computationally too expensive to be solved in practice, and it is justified by the testings.

Therefore, in this project, we present and compare three different methods from ML solution to estimate the parameters of the \textsc{Poisson-Gaussian noise}. The first method is based on the variance of the
noisy image for each pixel value of the real image. Then, we will present a
method based on the cumulant of the noisy image and the knowledge we have of the
real image. Finally, we implemented a basic convolution neural network in order
to compare our result.

\subsection{\textsc{Variance}}
This method is based on the variance of the values of the noisy image for a
fixed intensity of the real image. That is to say, we take a pixel $i$ from $x$
of intensity $x_i$, then for every $j$ such that $x_j = x_i$, we have $y_j
    \simPG[x_i]$. Thus, if we denote $Y_i = \set{y_k : x_k = x_i}$, we have
$\mathbb{V} [Y_i] \approx \frac{x_i}{a} + b^2$. We can calculate this variance
with each distinct value of $x_i$. In our case, images are saved in 8-bits, thus
we only have 256 unique different values of $x_i$. Moreover, as we know the
theoretical mean of $Y_i$ is $x_i$, we can calculate the variance using :

\begin{equation}
    \mathbb{V} [Y_i] = \frac{1}{\card{Y_i}} \sum_{y_k \in Y_i} (y_k - x_i)^2
\end{equation}

Finally, to obtain the estimation of $a,b$ is :
\begin{equation}
    \hat a, \hat b = \arg \min_{a,b} {\sum_i (\mathbb{V} [Y_i] - \frac{x_i}{a} - b^2)^2}
    \label{eq:var}
\end{equation}

Note that in equation~\ref{eq:var}, the same point $(x_i,\mathbb{V} [Y_i])$ is
present $\card{Y_i}$ times. This is because we found better result using this
bias. This method has multiple default, first it is not unbiased, then it works
best on images with a small amount of unique pixel intensities but an important
difference between the minimum and maximum intensity. Also, because it is
biased, this method can be tuned to be better (for instance, the importance of
each terms on the right side of equation~\ref{eq:var} can be modified so that
higher values of $x_i$ has a smaller weight).

\subsection{\textsc{Cumulant}}
This method uses the cumulant expansion of the noisy image. In this section,
instead of seeing $x$ and $y$ as images, we see $x$ and $y$ as samples from a
distribution where $x \sim \mathcal{X}$ and $y \sim \mathcal{Y}$ such that :
\begin{equation*}
    \mathbb{P} [x = x_i] = \frac{\card{\set{k : x_k = x_i}}}{n}
\end{equation*}
where $n$ correspond to the number of sample (i.e., the size of $x$ and $y$). Then
we can define $\mathcal{Y}$ as the distribution of \textsc{Poisson-Gaussian noise} over the
distribution $\mathcal{X}$. Formally:
\begin{equation}
    \mathcal{Y} \sim \frac{\mathcal{P}(a \mathcal{X})}{a} + \mathcal{N}(0,b^2)
\end{equation}
We then use the equation~\ref{eq:cumu} calculated in
appendix~\ref{appendix:cumu} to get the cumulant of $\mathcal{Y}$ as a system
of two equations :
\begin{equation}
    \begin{split}
        \kappa_2 &= \frac{\overline{x}}{a} + \overline{x^2} - \overline{x}^2 + b^2 \\
        \kappa_3 &= \overline{x^3} - 3 \overline{x^2}\overline{x} + 2 \overline{x}^3 + 3\frac{\overline{x^2}}{a} - 3 \frac{\overline{x}^2}{a} + \frac{\overline{x}}{a^2}
    \end{split}
    \label{eq:cumu}
\end{equation}
where $\overline{x^k}^j = (\frac{1}{n}\sum_i x_i^k)^j$. Equations~\ref{eq:cumu} forms
a system of two equations with two variables : $a$ and $b$, the parameters of
the noise. This method benefits being unbiased for finding $\kappa_{2,3}$, some
extra-calculation can be made so make $\hat a$ and $\hat b$ unbiased.

\subsection{\textsc{Cnn}}
\label{sec:cnn}
For the sake of comparison with our methods presented above, we implement a convolutional regression network trained to predict $a$ and $b$. It uses fairly standard layers, but isn't inspired by any particular architecture. For optimization, we used an Adam~\cite{kingma2017adam} optimizer and for the loss we picked Mean Absolute Percentage Error, which is given by
\begin{equation}
    \frac{100}{N}\sum_{n=1}^{N}\left| \frac{v_{pred,n} - v_{real,n}}{v_{real,n}} \right|
\end{equation}
where $v_{real,n}$ are the predictions made by the model and $v_{real,n}$ the ground-truth values. This specific loss is nice because it is normalized by the real value, hence errors for big values are not over penalized.

The detailed architecture of the \textsc{Cnn} can be found in table~\ref{tab:cnn_architecure}.

\begin{table}[ht]
    \caption{Architecture of the \textsc{Cnn}}
    \centering
    \begin{tabular}{c V{4} c  c}
        Layer     & Out channels & Parameters                                               \\  \thickhline
        Input     & $1$          & -                                                        \\
        Conv2D    & $16$         & $\text{kernel\_size}=(3, 3)\text{, padding}=\text{same}$ \\
        ReLU      & $16$         & -                                                        \\
        BatchNorm & $16$         & $\text{over the channels}$                               \\
        MaxPool2D & $16$         & $\text{pool\_size}=(2, 2)$                               \\
        Conv2D    & $32$         & $\text{kernel\_size}=(3, 3)\text{, padding}=\text{same}$ \\
        ReLU      & $32$         & -                                                        \\
        BatchNorm & $32$         & $\text{over the channels}$                               \\
        MaxPool2D & $32$         & $\text{pool\_size}=(2, 2)$                               \\
        Conv2D    & $64$         & $\text{kernel\_size}=(3, 3)\text{, padding}=\text{same}$ \\
        ReLU      & $64$         & -                                                        \\
        BatchNorm & $64$         & $\text{over the channels}$                               \\
        MaxPool2D & $64$         & $\text{pool\_size}=(2, 2)$                               \\
        Dense     & $16$         & -                                                        \\
        ReLU      & $16$         & -                                                        \\
        BatchNorm & $16$         & $\text{over the channels}$                               \\
        Dropout   & $16$         & $\text{rate}=0.5$                                        \\
        Dense     & $4$          & -                                                        \\
        ReLU      & $4$          & -                                                        \\
        Dense     & $2$          & -                                                        \\
        Linear    & $2$          & -                                                        \\
    \end{tabular}
    \label{tab:cnn_architecure}
\end{table}

We tested different version of inputs to the network: $x$ only, $x$ and $y$ concatenated, $x$, $y$ and $|y-x|$ (rescaled to $[0., 1.]$) concatenated and for each of these three, another version which is shifted to the range $[-.5,.5]$. For training and testing, we used a dataset called Berkeley Segmentation Dataset 300~\cite{dataset:MartinFTM01}. For each sample, we draw uniformly at random $a$ and $b$ in some predefined ranges, and then synthesize a $y$ for generating the respective input to the network.

\subsection{\textsc{Foi et al.}}
The method \textsc{Foi et al.}~\cite{foi2008practical} estimates $a$ and $b$ using a preprocessing step where the wavelet transform is applied on the image and then the pixels segmented into non-overlapping intensity level sets. Further, a some local estimation of multiple expectation/standard-deviation pairs is performed before finally a global parametric model fitting to those local estimates is done. We only implemented an interface, or wrapper, for this code to be integrable with our code. Other than that, we used it as is. But it's important to note that the authors of~\cite{foi2008practical} use a slightly different noise model, where their $a$ and $b$ are both inversely proportional to our $a$ and $b$.

\section{Results and Discussion}

In this section, we will describe the performance of the different estimation
methods. But first, we need to quantify the quality of an estimation $\hat
    a,\hat b$ of $a,b$. Since in equations~\ref{eq:var}~and~\ref{eq:cumu}, we have a
polynomial of $\frac{1}{a}$ and $b^2$, our goal will be on one side to minimize
$(\frac{1}{\hat{a}} - \frac{1}{a})^2$ and on the other to minimize $(\hat{b}^2
    -b^2)^2$. From now on, saying the estimation of $a$ refers in fact to the
estimation of $\frac{1}{a}$ such as depicted before, same thing for $b$ with
$b^2$.

\subsection{Dataset}

Our evaluation data is based on the Berkeley Segmentation
Dataset 300~\cite{dataset:MartinFTM01}. We use $10$ of those images that we considered
as ground-truth, we then added the same noise with $10$ different seeds, making
$100$ images to have estimation on for every $a,b$. We then used $25$ linearly
spaced values of $a$ and $b$ with $a \in [1., 100.]$ and $b \in [.01,.15]$.
Making a total of $62500$ estimations for each method.

\subsection{Benchmarking methods}

\subsubsection{\textsc{Cnn}}
As mentioned in section~\ref{sec:cnn}, we have tried different modes of the \textsc{Cnn} all having different types of inputs. Testing all those against each other lead to surprising results. Interestingly, the version using only the non-shifted $x$, basically replicating the setting of \textsc{Foi et al.}, did perform best. But, this has more to do with the expressiveness of our model than the general ability of \textsc{Cnn} to predict the noise parameters. Having more information at hand, for example $y$ and $x$, should ultimately lead to better performance. So, a more powerful model having a deeper architecture would be an important point to investigate further. For the remainder of the report, we shall use \textsc{CNN\_N} to describe the \textsc{Cnn} method using only $x$ (\textsc{N} for \textit{noisy}) without shifting as inputs.

\subsubsection{\textsc{Foi et al.}}
The results we got from \textsc{Foi et al.} compared to the other methods are generally quite bad, which came as a surprise for us. On one hand, a reasonable explanation for this is the fact that in \textsc{Foi et al.} only $y$ is used but on the other hand \textsc{CNN\_N} also only uses $y$ but still performs better. For most of the following graphs and plots, we only focus on the other three methods because of the underwhelming performance of \textsc{Foi et al.} in our setting.

\subsection{Overall scores}

\begin{table}
    \caption{Statistics about the error on $a$ for various methods}
    \centering
    \begin{tabular}{c V{4} c ccc}
        Method & mean      & std       & 75\%-quantile & max        \\ \thickhline
        CUMU   & $.000003$ & $.000014$ & $.000001$     & $.000277$  \\
        VAR0   & $.000008$ & $.000085$ & $.000000$     & $.003535$  \\
        CNN\_N & $.017767$ & $.086703$ & $.000074$     & $.633754$  \\
        FOI    & $3147$    & $746292$  & $.000564$     & $186.10^6$ \\
    \end{tabular}
    \label{tab:stat_a}
\end{table}

\begin{table}[ht]
    \caption{Statistics about the error on $b$ for various methods}
    \centering
    \begin{tabular}{c V{4} c ccc}z
        Method & mean      & std        & 75\%-quantile & max          \\ \thickhline
        CUMU   & $.000000$ & $.000001$  & $.000000$     & $.000033$    \\
        VAR0   & $.000001$ & $.000011$  & $.000000$     & $.000445$    \\
        CNN\_N & $.000008$ & $.000023$  & $.000005$     & $.000387$    \\
        FOI    & $.346023$ & $6.674889$ & $.000094$     & $615.874927$ \\
    \end{tabular}
    \label{tab:stat_b}
\end{table}

Looking at the statistics shown in Table~\ref{tab:stat_a} and
Table~\ref{tab:stat_b} reveals that both \textsc{Variance} and \textsc{Cumulant} performs better than the
\textsc{Cnn} and \textsc{Foi et al.}. Looking at the \textsc{mean} or the \textsc{max} columns shows
overall better result for \textsc{Cumulant} than for \textsc{Variance}. However, those results are to be
taken with a pinch of salt.

\subsection{Bias}

\subsubsection{Dependence on \texorpdfstring{$a$}{a} or \texorpdfstring{$b$}{b}}
\label{section:bias_dep_a_b}

\begin{figure}[ht]
    \centering
    \includegraphics[width = \columnwidth]{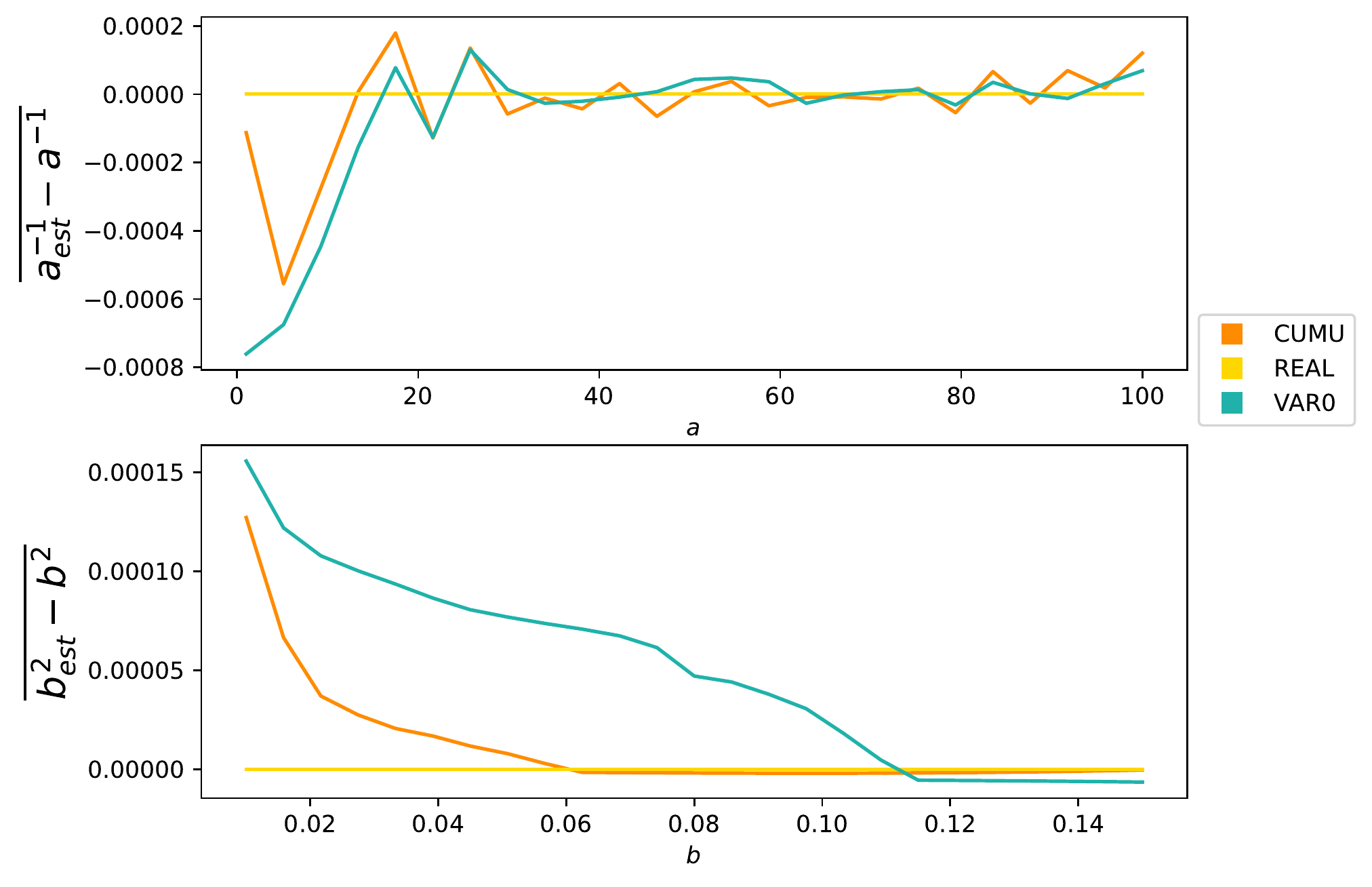}
    \caption{Bias of $a_{est}$ and $b_{est}$ for \textsc{Variance} and \textsc{Cumulant}.}
    \label{fig:biasness_3_methods}
\end{figure}

The comparison method used in figure~\ref{fig:biasness_3_methods} comes from
the fact that both equations~\ref{eq:var}~and~\ref{eq:cumu} are polynomial of
$\frac{1}{a}$ and $b^2$. Thus, because we used an unbiased estimator for $\kappa_2$
and $\kappa_3$, \textsc{Cumulant} is nearly unbiased with respect to this formula when $a$ is large enough that $\frac{1}{a^2}$ is small in the formula of $\kappa_3$. However,
we see from figure~\ref{fig:biasness_3_methods} that it is indeed the case for
high values of $a$ and $b$ but not for low values of $a$ and $b$. This can be
explained by the fact that we only kept \textit{realistic} values of $b$.
Indeed, if we had kept every estimation of $b^2$, we would have had a
mean of $0$. Since we only kept values of $b^2$ that are higher than $0$, we
have a positive bias. Equation~\ref{eq:var} shows that under-estimating $b$
leads to an over-estimation of $1/a$ which makes assuming $b^2 = 0$ and re-do
calculation of $a$ based on that a way to help minimize bias. Although bias is
not that important because for small values of $b$ it results a misestimation
of $b^2$ and not biasedness of our method. Figure~\ref{fig:biasness_3_methods}
also shows that \textsc{Variance} is always biased for $b$ while \textsc{Cumulant} seems to settle down
at $b \approx .06$.

Looking at bias as a function of both $a$ and $b$ and not only $a$ or $b$ gives
interesting insights on the way \textsc{Variance} and \textsc{Cumulant} estimates $a$ and $b$.

\begin{figure}[ht]
    \includegraphics[width = \columnwidth]{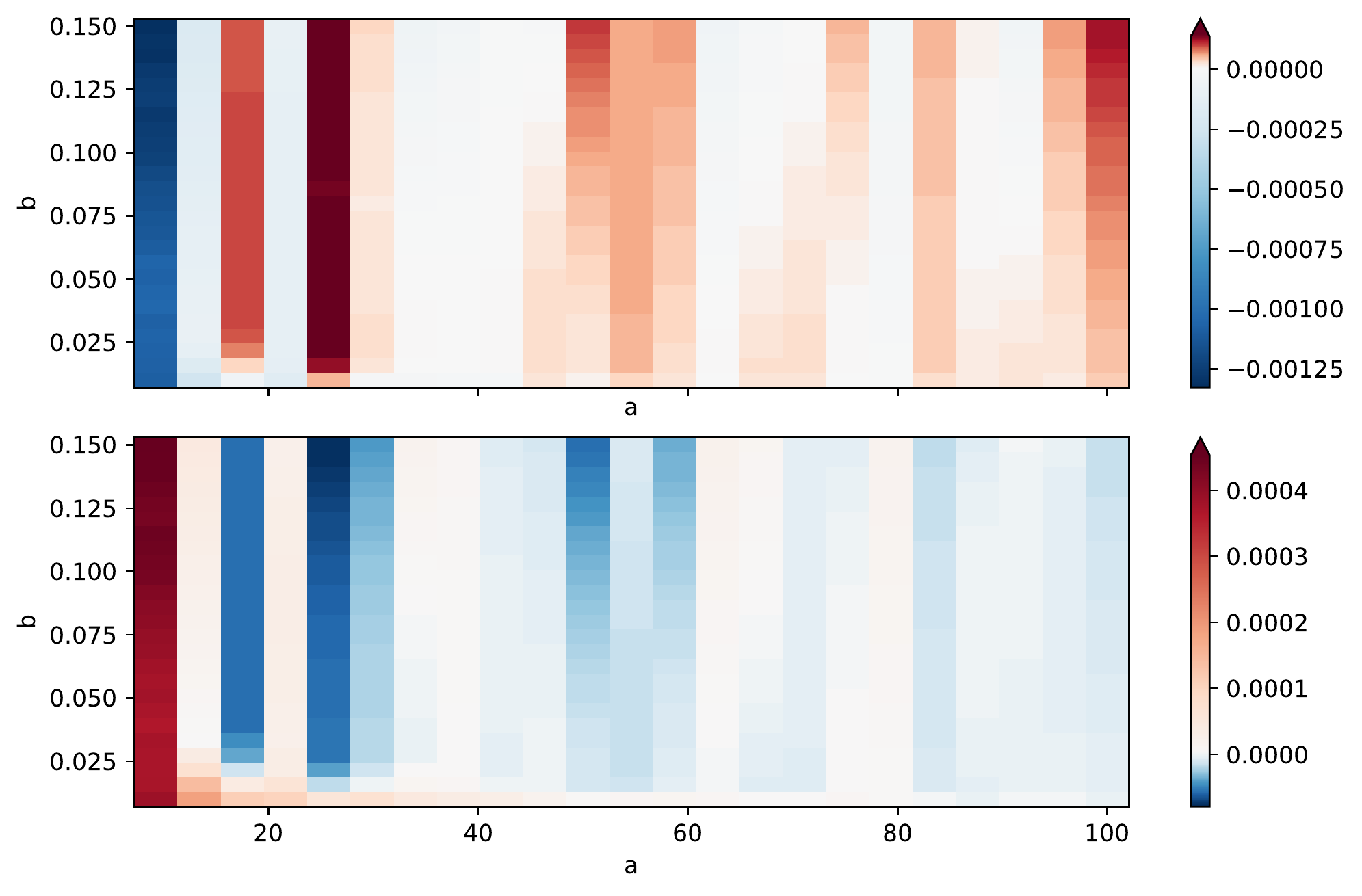}
    \caption{Bias of $a_{est}$ and $b_{est}$ for \textsc{Variance}.}
    \label{fig:bias_map_var0}
\end{figure}

\subsubsection{Bias as a function of \texorpdfstring{$a$}{a} and \texorpdfstring{$b$}{b}}

For readability, in this section, we removed the columns corresponding to the two
smallest values of $a$.

\begin{figure}[ht]
    \includegraphics[width = \columnwidth]{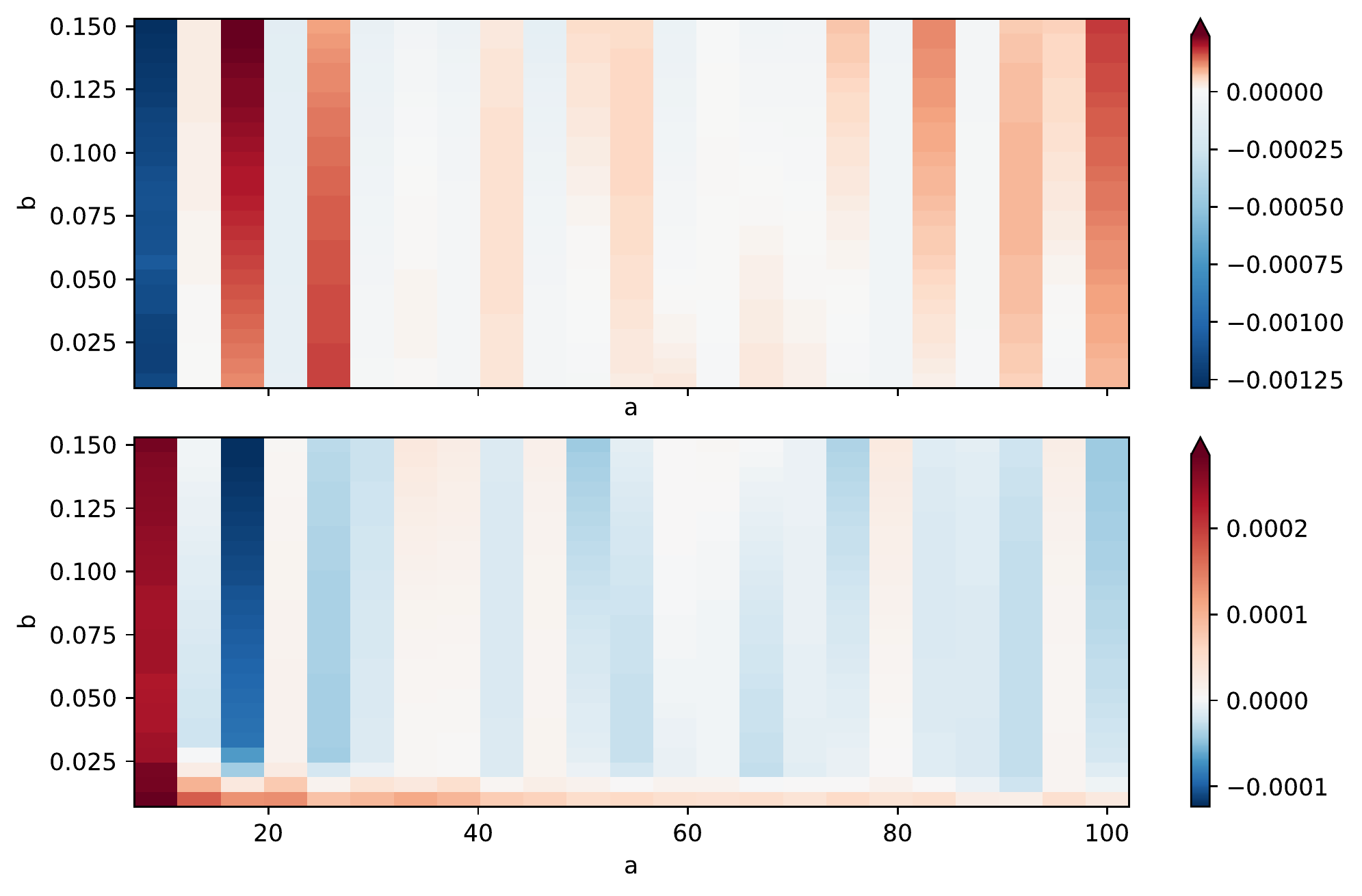}
    \caption{Bias of $a_{est}$ and $b_{est}$ for \textsc{Cumulant}.}
    \label{fig:bias_map_cumu}
\end{figure}

Figure~\ref{fig:bias_map_var0} and figure~\ref{fig:bias_map_cumu} are picturing columns that show the bias on $a_{est}$ or $b_{est}$ as a function of $a_{real}$ and $b_{real}$. In both figures, we see a high dependence from the sign of the bias on $a_{real}$. We do not have an exact explanation on that, but it may be a result of implementation of the noise generator more than a ground truth for the methods\footnote{And if not, good news : because the bias is dependent on $b$, if we just add Gaussian noise, we may be able to interpolate the value of our bias on $b_{est}$ and then find a better $b_{est}$}. This hypothesis also lean on the fact that \textsc{Cumulant} should be unbiased, and figure~\ref{fig:bias_map_cumu} happens to have the same bias on the same columns as figure~\ref{fig:bias_map_var0}. We can also see that whenever a column in biased toward positive for $a$ then it is biased for $b$ but the other way around, which highlights the confusion of the noise parameters. Even if it may look like \textsc{Variance} and \textsc{Cumulant} are biased the same way, dry results~\ref{tab:bias_var_cumu} shows that this is not the case.

\begin{table}[ht]
    \centering
    \caption{Bias of \textsc{Variance} and \textsc{Cumulant}}
    \label{tab:bias_var_cumu}
    \begin{tabular}{c V{4} c | c}
                      & $a$   & $b$   \\ \thickhline
        Same bias     & 29290 & 28824 \\
        Opposite bias & 27261 & 26550
    \end{tabular}
\end{table}

\begin{figure}[ht]
    \centering
    \includegraphics[scale=0.45]{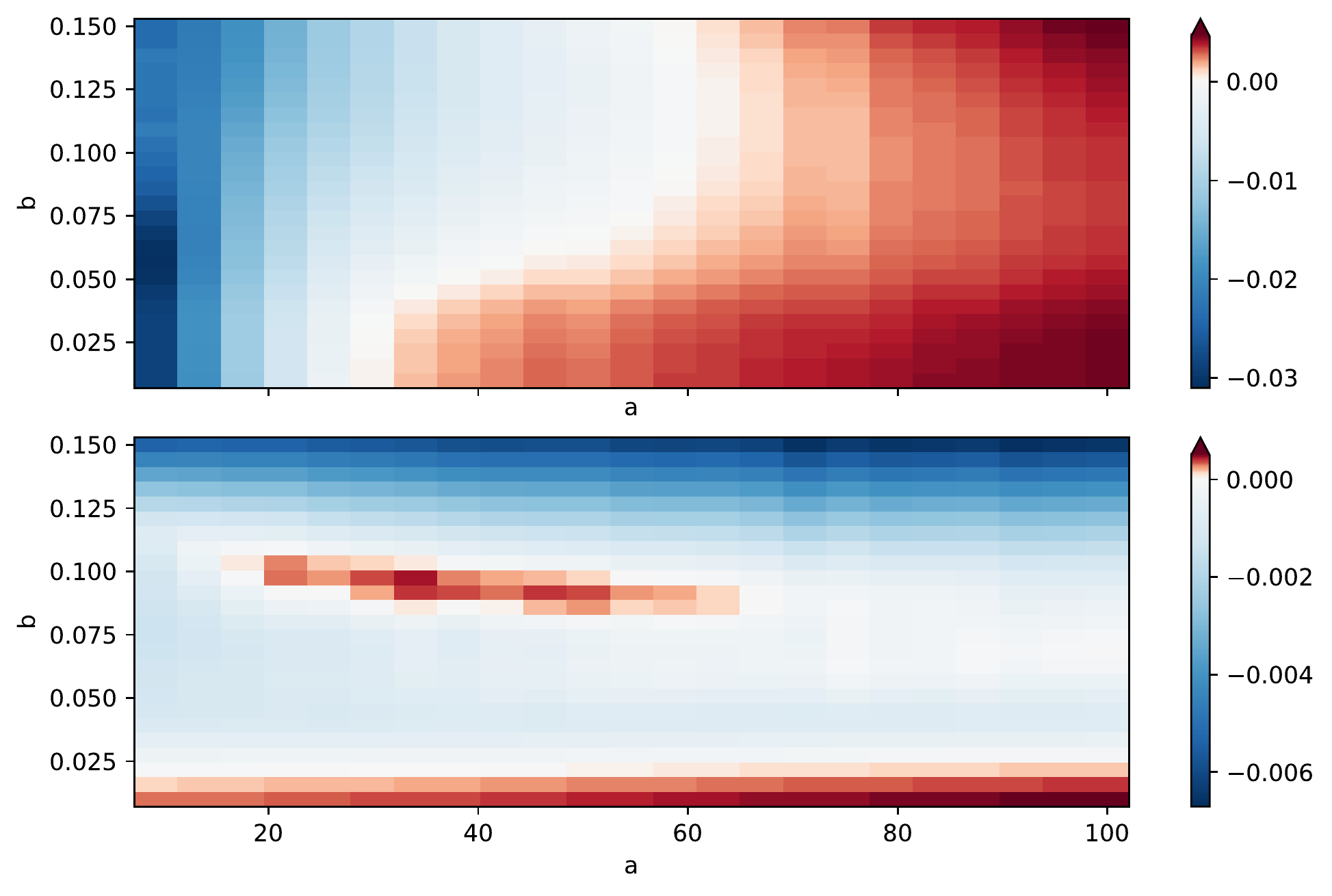}
    \caption{Bias of $a_{est}$ and $b_{est}$ for \textsc{Cnn}.}
    \label{fig:bias_map_cnn}
\end{figure}

\begin{figure}[ht]
    \centering
    \includegraphics[scale=0.45]{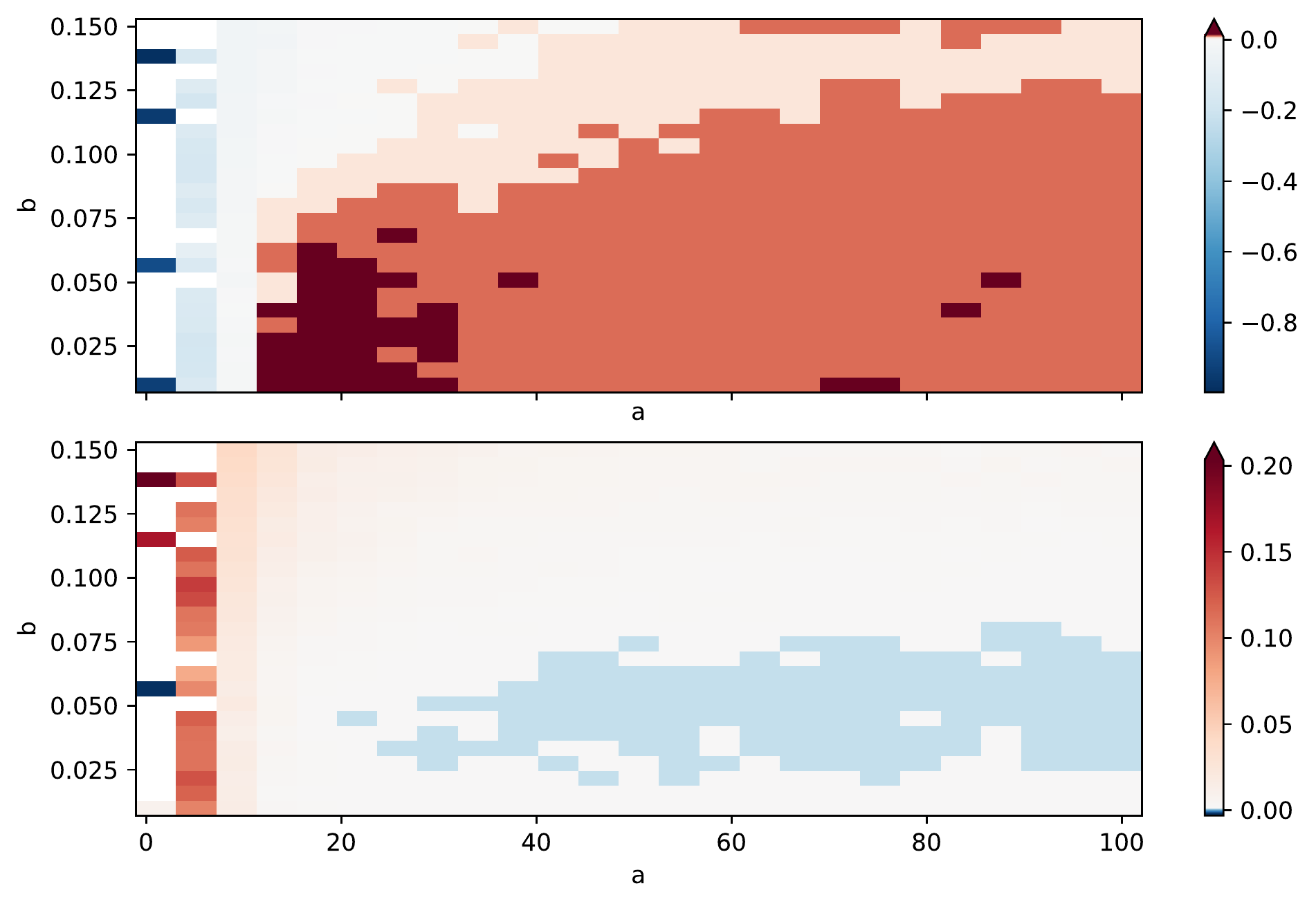}
    \caption{Bias of $a_{est}$ and $b_{est}$ for \textsc{Foi et al.}.}
    \label{fig:bias_map_foi}
\end{figure}

Moreover, figure~\ref{fig:bias_map_cnn} shows bias toward the extrema of $a$ for
the bias of $a$ and $b$ for the bias of $b$. This can be caused by the \textsc{Cnn}
being bad on the limits of its training set. However, \textsc{Cnn} does not seem to
recognize the notion of noise as there is no dependency of the bias on $a$ with
the bias on $b$.

Looking at figure~\ref{fig:bias_map_foi} shows that \textsc{Foi et al.}
positive bias on $a$ is explained by negative bias on $b$ and vice-versa. This
shows that \textsc{Cnn} could learn this behavior as \textsc{Foi et al.} and \textsc{Cnn} take the same input.

In section~\ref{section:bias_dep_a_b} we said the bias is caused by the removal
of unrealistic values of $b_{est}$. In the next section, we will analyze up to
what extent this is true.

\subsubsection{Realistic estimations}
\begin{figure}[ht]
    \centering
    \includegraphics[scale=0.55]{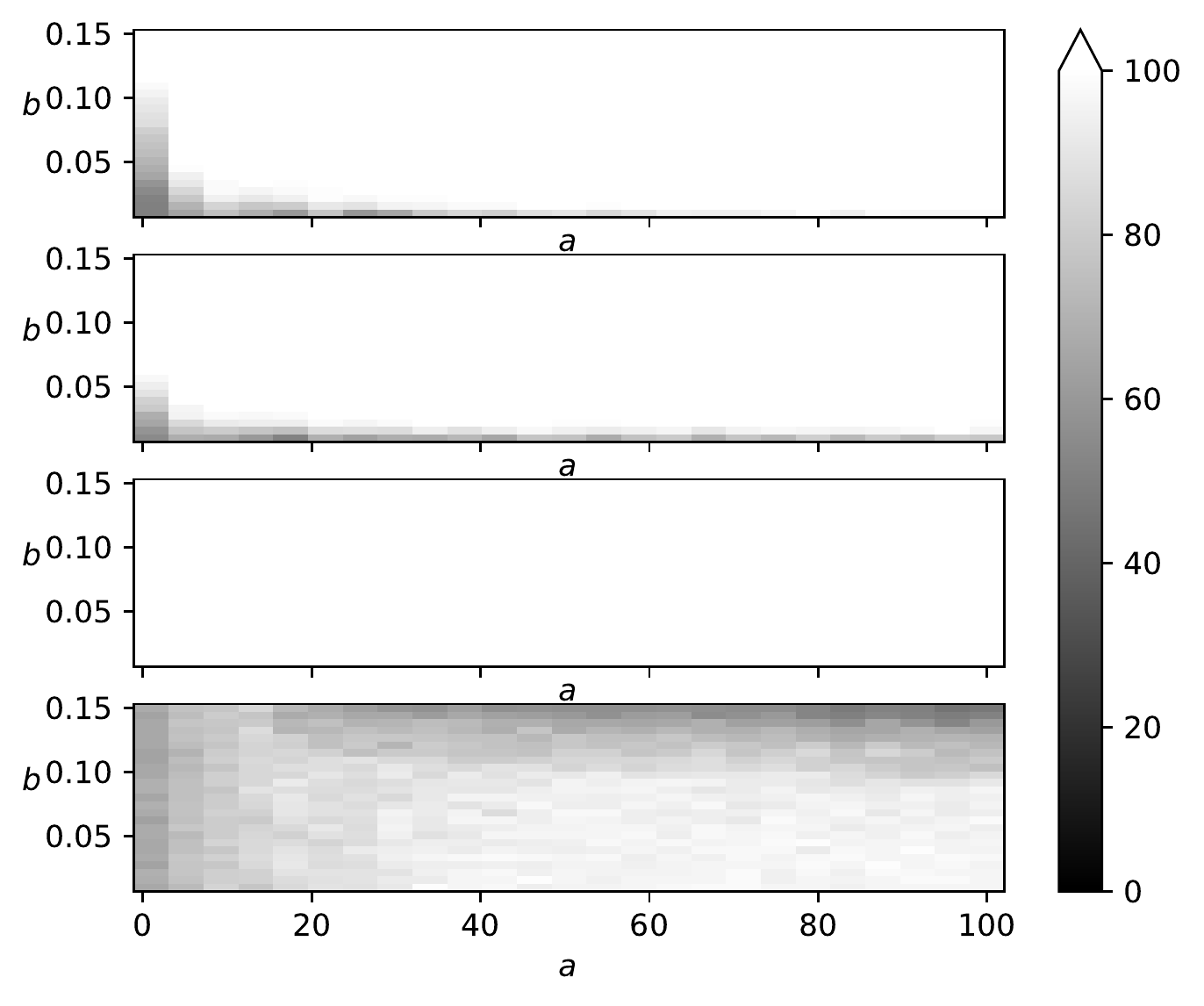}
    \caption{Percentage of realistic $b_{est}$ for each method.}
    \label{fig:percentage_b_is_found}
\end{figure}

\begin{figure}[ht]
    \includegraphics[width = \columnwidth]{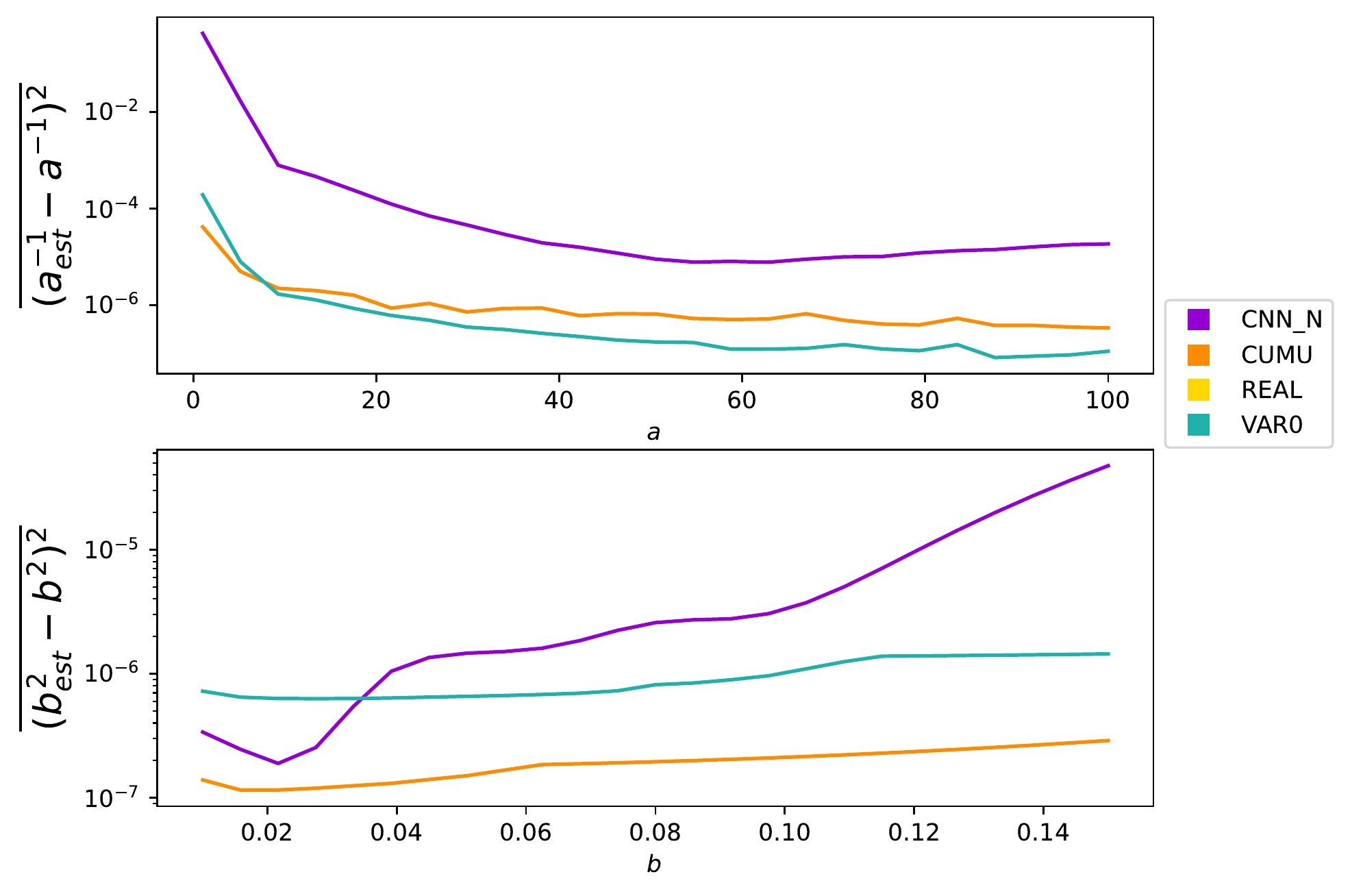}
    \caption{Mean squared error for each method as a function of $a$ (top) and $b$ (bottom).}
    \label{fig:MSE_each_method}
\end{figure}

It is found that every method is finding
realistic $a$ (except \textsc{Foi et al.}). This does not sound absurd as our range of $a$ (up
to $a = 100$) allowed misestimation of $\frac{1}{a}$ up to $10^{-2}$ which is
greater than what we found with \textsc{Variance} and \textsc{Cumulant} in
figure~\ref{fig:MSE_each_method}.

For \textsc{Cnn} the reason is different, the architecture forced the values of $a_{est}$ and $b_{est}$ to be inside the range of the training set. Interesting result about \textsc{Foi et al.} is that $a_{est}$ was realistic if and only if $b_{est}$ was. This is not the case for the other methods, as shown in figure~\ref{fig:percentage_b_is_found}. \textsc{Cnn} finds every $b_{est}$ again by construction. However, we can see noticeable difference between \textsc{Variance} and \textsc{Cumulant} when it comes to the proportion of realistic $b_{est}$
found.

\begin{figure}[ht]
    \centering
    \includegraphics[scale=0.45]{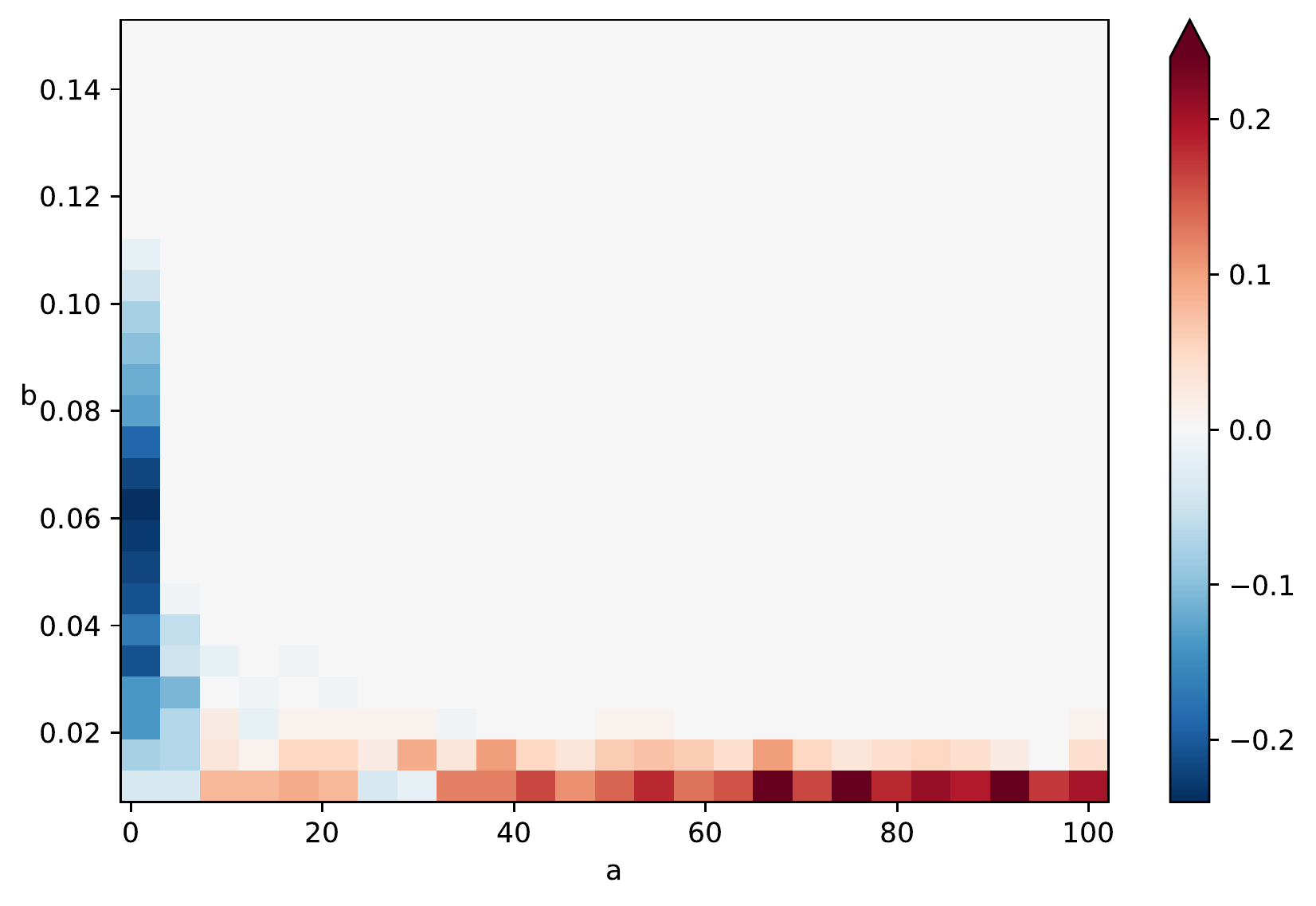}
    \caption{Relative quantity of realistic $b_{est}$, the redder, the more \textsc{Variance}
        found $b_{est}$ compared to \textsc{Cumulant}}
    \label{fig:percentage_b_is_found_var_cumu}
\end{figure}

Figure~\ref{fig:percentage_b_is_found_var_cumu} highlights the difficulties of \textsc{Variance} to realistically estimate $b$ when it is small, while \textsc{Cumulant} seems to have more difficulties when it comes to small $a$. For \textsc{Cumulant}, this can be explained by the variance on $\kappa_3$ as this variance is dependent on $a^{-1}$ thus greater when $a$ is small. Since $a_{cumulant}$ only depends on $\kappa_3$, the precision of $a_{cumulant}$ depends on the precision of the estimation of $\kappa_3$ which depends directly on $a_{real}$.

\subsection{Mean squared error}

The MSE of $\hat a$ is dependent on $a$. We see on
Figure~\ref{fig:MSE_each_method} that the smaller $a$, the worst the estimation
is. This is a consequence of using $\frac{1}{a}$. When $a$ is small, the Poisson
noise is more present and when $a$ increases it is disappearing, all of our
methods seems to be better at recognizing small noise over a lot of noise even
though the MSE on $b$ does not seem to depend a lot on the value of $b$ for both
\textsc{Variance} and \textsc{Cumulant}.

We also see that both of the presented methods are better than
\textsc{Cnn} at finding result. We also see that \textsc{Cumulant} is 10 times better than \textsc{Variance} at
estimating $b$ for this error.

\subsection[ht]{Outliers}

When evaluating our methods, we realized that some outliers have a big influence on the overall performance. So, in this section, we show how they impact the estimations.

\begin{figure}[ht]
    \centering
    \includegraphics[scale=0.45]{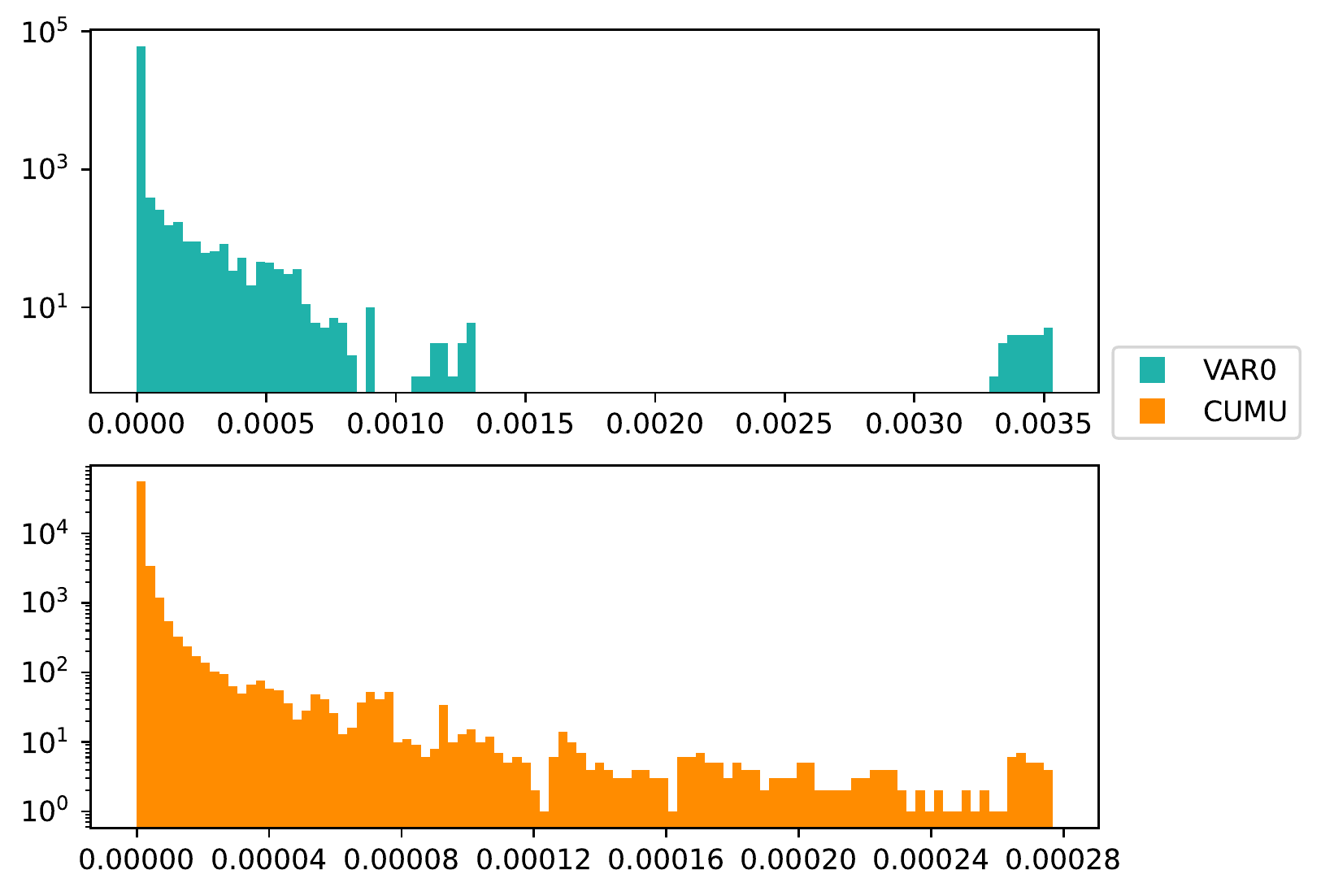}
    \caption{Distribution of $MSE(a)$ for \textsc{Variance} and \textsc{Cumulant}}
    \label{fig:mse_a_distribution}
\end{figure}

\begin{figure}[ht]
    \centering
    \includegraphics[scale=0.45]{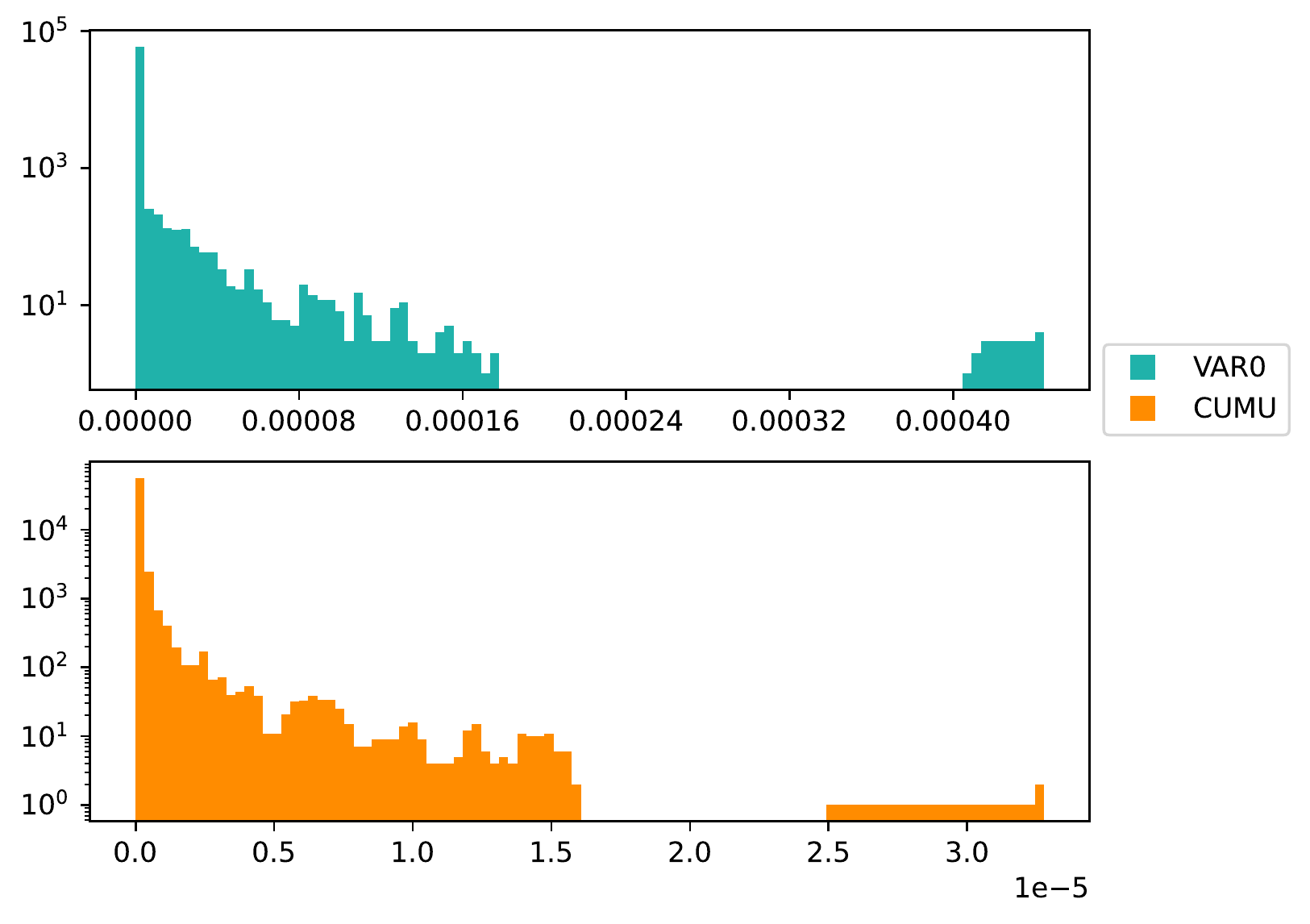}
    \caption{Distribution of $MSE(b)$ for \textsc{Variance} and \textsc{Cumulant}}
    \label{fig:mse_b_distribution}
\end{figure}

\begin{figure}[ht]
    \centering
    \includegraphics[scale=0.5]{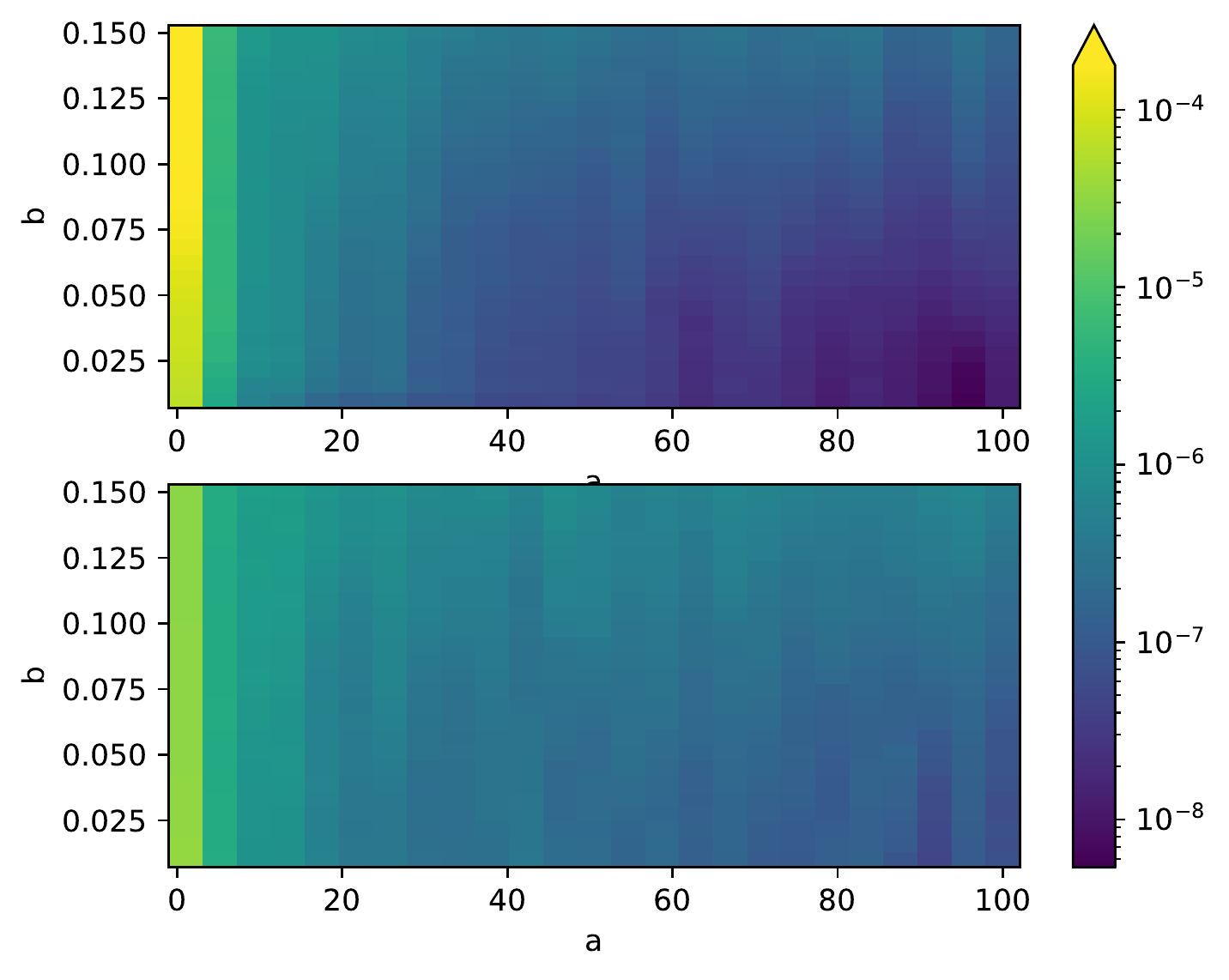}
    \caption{$MSE(a)$ of \textsc{Variance} and \textsc{Cumulant} after removing outliers}
    \label{fig:MSE_VAR_CUMU_a_wo_outliers}
\end{figure}

\begin{figure}[ht]
    \centering
    \includegraphics[scale=0.5]{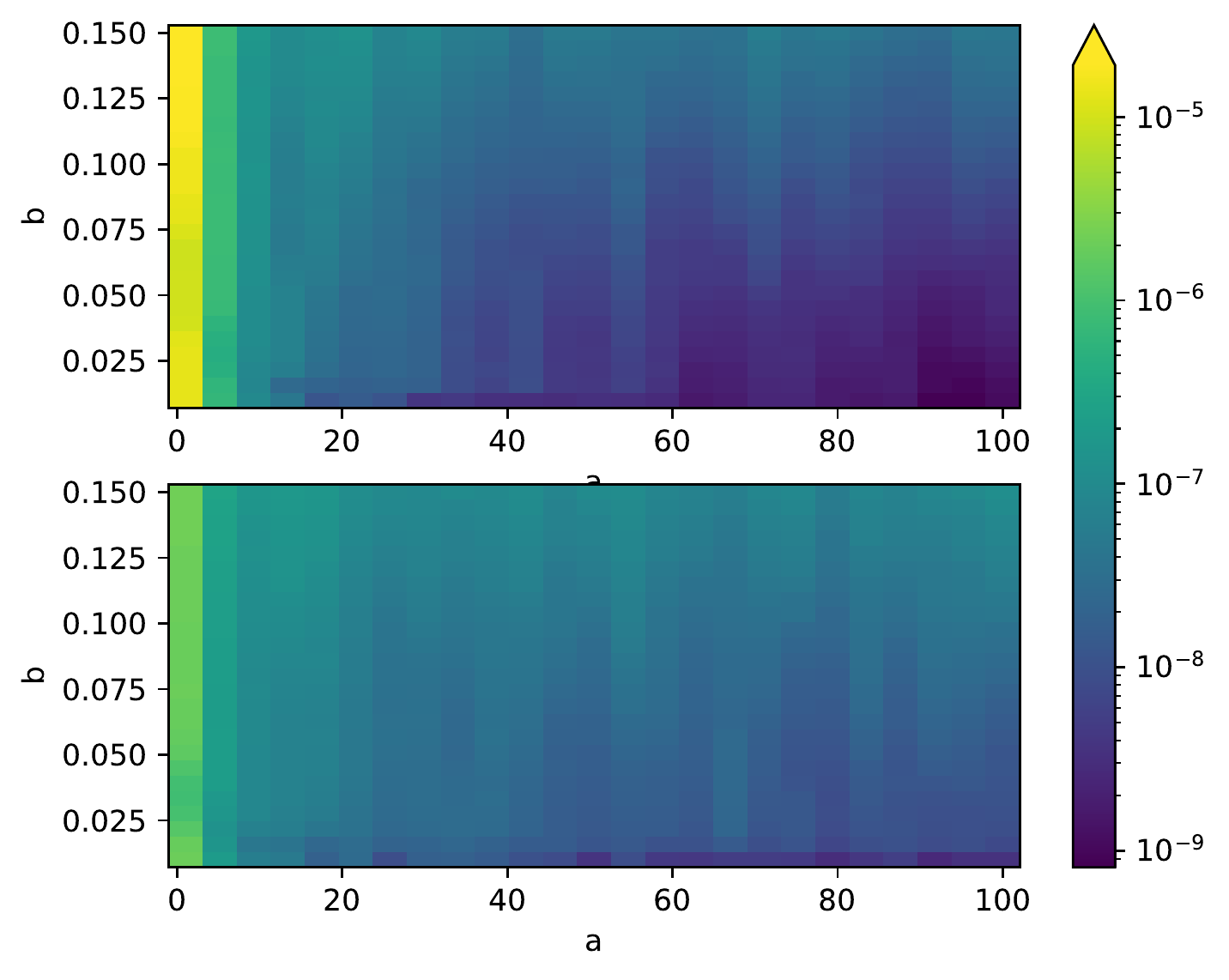}
    \caption{$MSE(b)$ of \textsc{Variance} and \textsc{Cumulant} after removing outliers}
    \label{fig:MSE_VAR_CUMU_b_wo_outliers}
\end{figure}

\subsection{Noise-free image dependence}

Here, we plot the dependence of the \textsc{MSE} on the $10$ round-truth image $x$ that was used to synthesize the noisy image $y$. We average over the $10$ different seeds and the $b$ or $a$ values respectively. One can observe that the ground-truth data, the image properties, have some influence on the estimation performance, more notably so on the estimation of $b$.

\begin{figure}[ht]
    \includegraphics[width = \columnwidth]{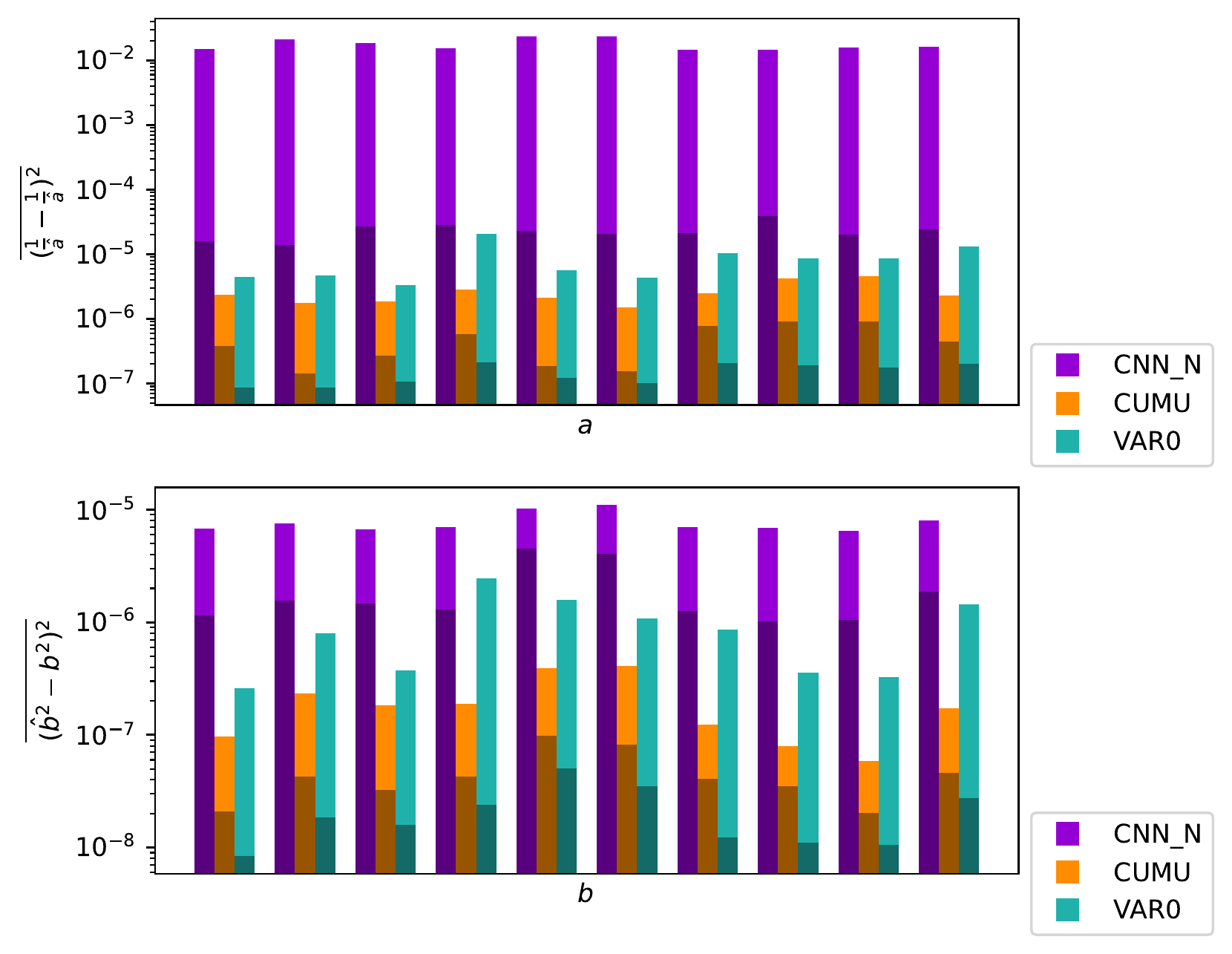}
    \caption{MSE depending on image for $a$ (top) and $b$ (bottom) with or without outliers (darkened)}
    \label{fig:MSE_dependence_image_outliers}
\end{figure}

\subsection{Log-likelihood}

For all the different data samples we have created for comparing the different methods with each other, we additionally computed their log-likelihood values. Those give another interesting perspective on the quality of those estimations.

\begin{figure}[ht]
    \centering
    \includegraphics[scale=0.5]{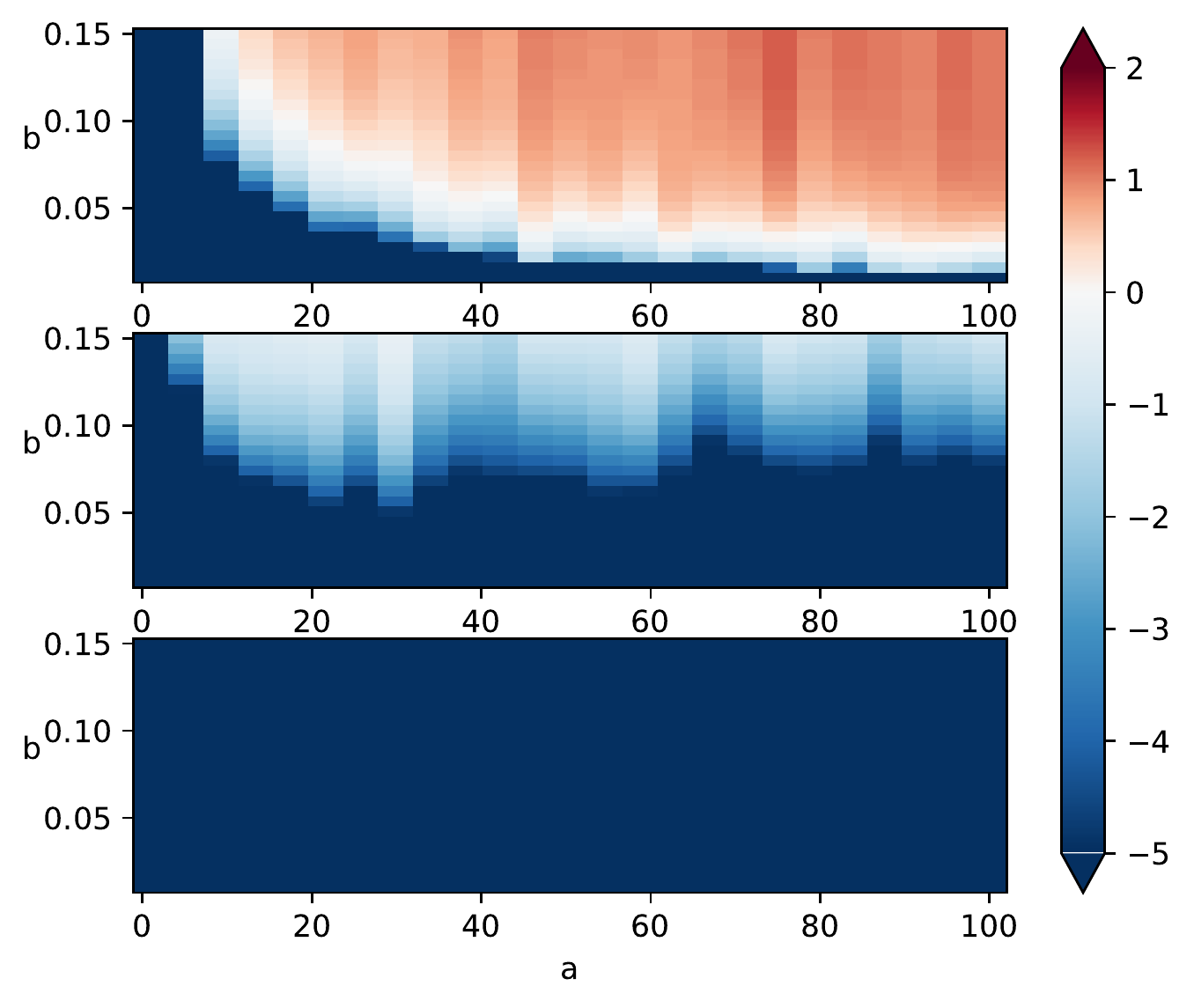}
    \caption{Difference $\mathcal{LL}$ of \textsc{Variance} and \textsc{Cumulant}}
    \label{fig:mean_dll}
\end{figure}

In terms of trying to maximize the likelihood, \textsc{Cnn} does not work as good as the other methods, as seen from Figure~\ref{fig:mean_dll}.

\section{Conclusion}

In this project, it's found and proven that Poisson-Gaussian noise modeling is properly matched with raw-data of digital sensors. In this modeling, noise can be decomposed into two parts: Poisson and Gaussian part, where the former one is signal dependent, while the latter is signal independent. We can also relate Poisson effect with discrete nature of photon counting process.
Based on this noise modelling, we defined two noise parameters and throughout the project, we tried to propose the algorithmic solution for this. Firstly, the likelihood function is derived as we also have the ground truth image as input. Then, to obtain the good estimation, we tried to obtain maximum likelihood solution of the noise parameter estimation. It's shown that it's practically inefficient to be implemented. Therefore, another types of solutions were proposed.
Two methods are proposed with using the statistical property of the noise modeling. The methods are \textsc{Variance} and \textsc{Cumulant}, where they use variance and cumulant information, respectively. \textsc{Cumulant} was found to be better at estimating values when $a$ was small, while \textsc{Variance} was found to be better when $b$ was small. For the real cases, Poisson part can dominate the noise, which means $a$ is small. Thus, we can conclude that \textsc{Cumulant} is more robust for real-world cases. Also, \textsc{Variance} relies on discrete intensities, which may not be realistic.

Also, to compare the proposed methods with the solution found in the literature, two more algorithms were tested: \textsc{Foi et al.} and \textsc{Cnn}. It's found that both \textsc{Variance} and \textsc{Cumulant} are better than these methods for MSE and likelihood comparison. Also, it's found that all these methods might be used as a starting point for maximization of likelihood for the future works.

Another future work might be listed as adjusting the weights for the \textsc{Variance} method, considering the clipping behavior of the images in real-world. Throughout the experiment, we didn't clip any images, which are not the case in the real world. For the last thing to do might be adding the efficient maximization of likelihood.

\appendix
\subsection{Derivation of Maximum Likelihood Solution}
\label{appendix:derivML}
\subsubsection{Poisson-Noise Modeling}
Let us denote observed noisy image as $y$ and ground-truth image as $x$. Then,
Poisson-Gaussian modelling can be explained from~\ref{eq:poisgaus}.
\begin{equation}
    y = \frac{1}{a} \alpha + \beta, \quad \alpha \sim \mathcal{P}(ax), \quad \beta \sim \mathcal{N}(0,b^2)
    \label{eq:poisgaus}
\end{equation}
When expectation of both sides are taken, the following equation~\ref{eq:exppoisgaus} is obtained with the use of linearity property of
expectation.
\begin{equation}
    \mathbb{E}[y] = \frac{1}{a}\mathbb{E}[\alpha]=\frac{1}{a}ax=x
    \label{eq:exppoisgaus}
\end{equation}
When variance is applied to both sides in~\ref{eq:poisgaus}, the following
equation~\ref{eq:varpoisgaus} is obtained.
\begin{equation}
    \mathbb{V}[y] = \mathbb{E}\left[\left(\frac{1}{a}\alpha + \beta\right)^2\right] - x^2 = \frac{1}{a^2}\mathbb{E}[\alpha^2] + b^2 - x^2
    \label{eq:varpoisgaus}
\end{equation}

Given $\mathbb{E}[\alpha^2]=ax+a^2x^2 $, we have:
\begin{equation}
    \mathbb{V}[y] = \frac{x}{a}+x^2+b^2-x^2 = \frac{x}{a} + b^2
\end{equation}

\subsubsection{Likelihood Function of Single-Pixel Image}
From the definition of the probability mass function of a Poisson random variable,
the following equation~\ref{eq:pmfalpha} is obtained.
\begin{equation}
    \mathbb{P}(\{\alpha=k\})= \frac{e^{-ax}(ax)^k}{k!}, \quad k\geq 0
    \label{eq:pmfalpha}
\end{equation}

From the relation between probability density function (PDF) and probability
mass function (PMF) of discrete random variable with the use of Dirac delta
function, i.e. $f_X(t)=\sum_{k\in \mathbb{Z}} \mathbb{P}(\{X=k\})\delta(t-k)$, we have:

\begin{equation}
    f_\alpha(t|a,x) = \sum_{k=0}^\infty \frac{e^{-ax}(ax)^k}{k!} \delta(t-k)
\end{equation}

Let us define $\alpha'=\frac{1}{a}\alpha$. Then, the cumulative distribution
function (CDF) of this random variable $\alpha'$ can be found as following:

\begin{equation}
    F_{\alpha'}(t) = \mathbb{P}(\{\alpha'\leq t\}) = \mathbb{P}(\{\alpha \leq at\}) = F_\alpha(at)
    \label{eq:alpha_de}
\end{equation}

When taking the derivative of equation~\ref{eq:alpha_de} is taken, the PDF can be found as:

\begin{equation}
    f_{\alpha'}(t)=\frac{dF_{\alpha'}(t)}{dt} = \frac{dF_\alpha(at)}{dt} = a f_\alpha(at)
\end{equation}

As $\alpha$ and $x$ are given, the likelihood function of Poisson part can be
found as following:

\begin{equation}
    \begin{split}
        f_{\alpha'}(t|a,x) &= a\sum_{k=0}^\infty \frac{e^{-ax}(ax)^k}{k!} \underbrace{\delta(at-k)}_{=\frac{1}{a}\delta(t-\frac{k}{a})}\\
        &=\sum_{k=0}^\infty \frac{e^{-ax}(ax)^k}{k!}\delta(t-k/a)
    \end{split}
\end{equation}

The likelihood function of a Gaussian random variable with $0$ mean is as
following:

\begin{equation}
    f_\beta(t|b) = \frac{1}{b\sqrt{2\pi}}e^{-t^2/2b^2}
\end{equation}

Let us find the likelihood function of $y$. Since we know that $\alpha'$ and
$\beta$ are independent to each other, we have:

\begin{equation}
    \begin{split}
        f_y(y|a,b,x) &= (f_{\alpha'}*f_\beta)(y|a,b,x)\\
        &= \sum_{k=0}^\infty \frac{(ax)^k}{k!b\sqrt{2\pi}}\exp{\left(-ax-\frac{(y-k/a)^2}{2b^2}\right)}
    \end{split}
\end{equation}

\subsubsection{Maximum Likelihood Solution for Single-Pixel Image}
Thus, the maximum likelihood solution for a single-pixel image is as following:
\begin{equation}
    \begin{split}
        \hat{a}, \hat{b} &= \arg \max_{a, b} f_y(y|a,b,x) \\
        &= \arg \max_{a, b} \sum_{k=0}^\infty \frac{(ax)^k}{k!b\sqrt{2\pi}}\exp{\left(-ax-\frac{(y-k/a)^2}{2b^2}\right)}
    \end{split}
\end{equation}

\subsubsection{Likelihood Function of Multi-Pixel Image}

We can denote images as vectors of pixels, like $y_{n}$ and $x_{n}$ where $n \in
    \mathbb{N}$. Hence, we have the following:

\begin{equation}
    f_y(y_{n}|a,b,x) =\sum_{k=0}^\infty \frac{(ax_n)^k}{k!b\sqrt{2\pi}}\exp{\left(-ax_n-\frac{(y_n-k/a)^2}{2b^2}\right)}
\end{equation}

Given $x$, i.e., the vector that contains all $x_{n}$, it can be seen that
$y_{n}$ and $y_{n'}$ are independent, $\forall n \neq n'$. Therefore, we have:

\begin{equation}
    f_y(y|a,b,x) =\prod_{n}\sum_{k=0}^\infty \frac{(ax_n)^k}{k!b\sqrt{2\pi}}\exp{\left(-ax_n-\frac{(y_n-k/a)^2}{2b^2}\right)}
\end{equation}

\subsubsection{Maximum Likelihood Solution for Multi-Pixel Image}
Hence, we get the following maximization problem:

\begin{equation}
    \begin{split}
        \hat{a}, \hat{b} = \arg \max_{a, b} \prod_{n}\sum_{k=0}^\infty \frac{(ax_n)^k}{k!b\sqrt{2\pi}}\exp{\left(-ax_n-\frac{(y_n-k/a)^2}{2b^2}\right)}
    \end{split}
\end{equation}

Using the strict monotonicity of the logarithm, we can simplify the optimization
problem while not altering its results.

\begin{equation}
    \begin{split}
        \hat{a}, \hat{b} = & \arg \max_{a, b} \\
        & \sum_{n} \log{\left(\sum_{k=0}^\infty \frac{(ax_n)^k}{k!b\sqrt{2\pi}} \exp{\left(-ax_n-\frac{(y_n-k/a)^2}{2b^2}\right)}\right)}
    \end{split}
    \label{argmaxpoisgaus}
\end{equation}

In order to be computable, we limit the range of $k$ to a maximum value $k_{max}$
which has to be chosen big enough to get a good approximation.

\begin{equation}
    \begin{split}
        \hat{a}, \hat{b} \approx & \arg \max_{a, b} \\
        & \sum_{n} \log{\left(\sum_{k=0}^{k_{max}} \frac{(ax_n)^k}{k!b\sqrt{2\pi}} \exp{\left(-ax_n-\frac{(y_n-k/a)^2}{2b^2}\right)} \right)}
    \end{split}
\end{equation}

\subsection{Cumulant}
\label{appendix:cumu}
\subsubsection{Some properties}

\subsubsection{\texorpdfstring{$\kappa_r$}{Kappa r}}

For a random variable $X$ following the distribution $\mathcal{X}$, we consider the cumulant-generating function defined as :
\begin{equation*}
    K_{\mathcal{X}}(t) = \log(\mathbb{E} [e^{Xt}])
\end{equation*}
Then we define $\kappa_r $the $r^{th}$ cumulant of $\mathcal{X}$ as:
\begin{equation*}
    \kappa_{r} := K_{\mathcal{X}}^{(r)}(0)
\end{equation*}
with $K_{\mathcal{X}}^{(r)}(0)$ being the $r$-th derivative of $K_{\mathcal{X}}$ evaluated in $0$.

\subsubsection{Linearity}

The cumulant-generating function of a sum of independent distribution is the sum of their cumulant-generating function :
\begin{equation}
    \begin{split}
        K_{\mathcal{X} + \mathcal{Y}}(t) &= \log(\mathbb{E}(e^{(X+Y)t})) \\
        &= \log(\mathbb{E}[e^{Xt+Yt}]) \\
        &= \log(\mathbb{E}[e^{Xt} e^{Yt}]) \\
        &= \log(\mathbb{E}[e^{Xt}] \mathbb{E}[e^{Yt}]) \\
        &= \log(\mathbb{E}[e^{Xt}]) + \log(\mathbb{E}[e^{Yt}]) \\
        &=K_{\mathcal{X}}(t) + K_{\mathcal{Y}}(t)
    \end{split}
    \label{eq:cumu_lin}
\end{equation}

\subsubsection{Homogeneous of degree}
The $r^{th}$ cumulant is homogeneous of degree $r$ :
\begin{equation}
    \kappa_r(a \mathcal{X}) = a^r \kappa_r(\mathcal{X})
    \label{eq:cumu_homo}
\end{equation}

\subsubsection{Unbiased estimator}
For a vector $x$ obtained by sampling independently and identically $n$ times
from the law $\mathcal{X}$, \cite{website:wolframalpha_cumu} describes an
unbiased estimator of $\kappa_{2,3}$, the $r^{th}$ cumulant of $\mathcal{X}$
with :
\begin{equation*}
    \begin{split}
        \kappa_{2}(\mathcal{X}) = \frac{n}{n-1} m_{2}(x), \quad
        \kappa_{3}(\mathcal{X}) = \frac{ n^{2} } {(n-1) (n-2)} m_{3}(x)
    \end{split}
\end{equation*}
With $m_2$ the sample variance and $m_3$ the $3^{rd}$ sample central moment,
that can be calculated using the formulas taken
from~\cite{website:wolframalpha_moment} :
\begin{equation*}
    \begin{split}
        m_2(x) &= \frac{n-1}{n} \sum_i (x_i - \overline{x})^2 \\
        m_3(x) &= \frac{(n-1)(n-2)}{n^2} \sum_i (x_i - \overline{x})^3
    \end{split}
\end{equation*}

\subsubsection{Cumulant of Poisson-Gaussian Noise}
We have $\mathcal{Y} \simPG[\mathcal{X}]$, we want to have a
$\kappa_2(\mathcal{Y})$ and $\kappa_3(\mathcal{Y})$ as a function of $a$ and
$b$. First, we use equation~\ref{eq:cumu_lin} : $\kappa_r(\mathcal{Y}) =
    \kappa_r(\frac{\mathcal{P}(a \mathcal{X})}{a}) + \kappa_r(\mathcal{N}(0,b^2))$.

\subsubsection{Gaussian noise component}

The cumulant of $\mathcal{N}(0,b^2)$ are known :
\begin{equation}
    \begin{split}
        \kappa_2(\mathcal{N}(0,b^2)) &= b^2 \\
        \kappa_3(\mathcal{N}(0,b^2)) &= 0
    \end{split}
    \label{eq:cumu_normal}
\end{equation}

\subsubsection{Poisson noise component}

Instead of trying to find the cumulant of $\frac{\mathcal{P}(a
        \mathcal{X})}{a}$, we can use equation~\ref{eq:cumu_homo} and find the cumulant
of $\mathcal{Z} \sim \mathcal{P}(a\mathcal{X})$.
\begin{equation*}
    e^{K_{\mathcal{Z}}(t)} = \sum_k \mathbb{P}[Z = k]e^{tk}
\end{equation*}
Moreover, we know that :
\begin{equation*}
    \begin{split}
        \mathbb{P}[Z = k] &= \sum_i \mathbb{P}[X = x_i] \mathbb{P}[Z = k | X = i] \\
        &= \sum_i n_i \frac{(a x_i)^k e^{-ax_i}}{k!}
    \end{split}
\end{equation*}
where $n_i = \frac{\card{\set{j : x_j = x_i}}}{n}$, the proportion of
intensities equal to the one of $x_i$.

Thus :
\begin{equation*}
    \begin{split}
        e^{K_{\mathcal{Z}}(t)} &= \sum_k \mathbb{P}[Z = k]e^{tk} \\
        &= \sum_k \sum_i n_i \frac{(a x_i)^k e^{-ax_i}}{k!}  \exp(t)^k \\
        &= \sum_i n_i \frac{e^{-ax_i}}{exp(-ax_i e^t)} \sum_k \frac{(a x_i e^t)^k \exp(-ax_i e^t)}{k!} \\
        &= \sum_i n_i \exp(a x_i (e^t - 1))
    \end{split}
\end{equation*}

If we note : $f : t \rightarrow \sum_i n_i \exp(a x_i (e^t - 1))$ Then : $K_{\mathcal{Z}}(t) = \log(f(t))$
We have :
\begin{equation*}
    \begin{split}
        K_{\mathcal{Z}}(t) &= \log(f(t)) \\
        K_{\mathcal{Z}}^{1}(t) &= \frac{f^{(1)}(t)}{f(t)} \\
        K_{\mathcal{Z}}^{2}(t) &= \frac{f^{(2)}(t) f(t) - f^{(1)}(t)^2}{f(t)^2} \\
        K_{\mathcal{Z}}^{3}(t) &= \frac{f(t)[f(t) f^{(3)}(t) - 3 f^{(2)}(t) f^{(1)}(t)] + 2 f^{(1)}(t)^3}{f(t)^3} \\
        K_{\mathcal{Z}}^{4}(t) &= \frac{f^{(4)}(t)}{f(t)} - \frac{f^{(1)}(t) f^{(3)}(t)}{f(t)^2} \\&- 3 \frac{f^{(3)}(t) f^{(1)}(t) + f^{(2)}(t)^2}{f(t)^4} \\&+ 12 \frac{f^{(1)}(t)^2 f^{(2)}(t)}{f(t)^3} - 6 \frac{f^{(1)}(t)^4}{f(t)^4}\\
    \end{split}
\end{equation*}
With
\begin{equation*}
    \begin{split}
        f(0) &= 1\\
        f^{(1)}(0) &= a \overline{x}\\
        f^{(2)}(0) &= a \overline{x} + a^2 \overline{x^2}\\
        f^{(3)}(0) &= a \overline{x} + 3a^2 \overline{x^2} + 2 a^3 \overline{x^3}\\
        f^{(4)}(0) &= a^4 \overline{x^4} + 7 a^3 \overline{x^3} + 7 a^2 \overline{x^2} + a \overline{x} \\
    \end{split}
\end{equation*}
Thus :
\begin{equation*}
    \begin{split}
        K_{\mathcal{Z}}(0) &= 0 \\
        K_{\mathcal{Z}}^{1}(0) &= a \overline{x} \\
        K_{\mathcal{Z}}^{2}(0) &= a \overline{x}  + a^2 \overline{x^2} - a^2 \overline{x}^2\\
        K_{\mathcal{Z}}^{3}(0) &= a^3[\overline{x^3} - 3 \overline{x^2}\overline{x} + 2 \overline{x}^3] + a^2[3 \overline{x^2} - 3 \overline{x}^2] + a \overline{x}
    \end{split}
\end{equation*}

Now, using equation~\ref{eq:cumu_homo}, we have :
\begin{equation}
    \begin{split}
        \kappa_2\left(\frac{\mathcal{P}(a \mathcal{X})}{a}\right) &= \frac{\overline{x}}{a} + \overline{x^2} - \overline{x}^2 \\
        \kappa_3\left(\frac{\mathcal{P}(a \mathcal{X})}{a}\right) &= \overline{x^3} - 3 \overline{x^2}\overline{x} + 2 \overline{x}^3 + 3\frac{\overline{x^2}}{a} - 3 \frac{\overline{x}^2}{a} + \frac{\overline{x}}{a^2}
    \end{split}
    \label{eq:cumu_poisson}
\end{equation}

\subsubsection{Everything together}

By putting equations~\ref{eq:cumu_normal}~and~\ref{eq:cumu_poisson} together, we
have :
\begin{equation*}
    \begin{split}
        \kappa_2(\mathcal{Y}) &= \frac{\overline{x}}{a} + \overline{x^2}-\overline{x}^2 + b^2 \\
        \kappa_3(\mathcal{Y}) &= \overline{x^3}-3 \overline{x^2}\overline{x} + 2 \overline{x}^3 + 3\frac{\overline{x^2}}{a}-3 \frac{\overline{x}^2}{a} + \frac{\overline{x}}{a^2}
    \end{split}
\end{equation*}

\bibliographystyle{IEEEtran}
\bibliography{report_arxiv}
\end{document}